\documentclass[aps,twocolumn,showpacs,floatfix]{revtex4}
\usepackage{amssymb}
\usepackage{float}
\usepackage{latexsym}
\usepackage{epstopdf}
\usepackage{amsmath}
\usepackage[mathcal]{eucal}
\usepackage{epsfig}
\usepackage{dcolumn}
\usepackage{amsmath}
\usepackage{amsfonts}
\usepackage{bm}
\usepackage{color}

\newcommand{\be}{\begin{equation}}
\newcommand{\ee}{\end{equation}}
\newcommand{\bea}{\begin{eqnarray}}
\newcommand{\eea}{\end{eqnarray}}

\newcommand{\p}{\partial}
\newcommand{\s}{\sigma}

\newcommand{\la}{\langle}
\newcommand{\ra}{\rangle}
\newcommand{\rd}{\mbox{d}}
\newcommand{\ri}{\mbox{i}}
\newcommand{\re}{\mbox{e}}

\begin{document}

\title{Effects of thermal phase fluctuations in a 2D superconductor:
  an exact result for the spectral function.} 

\author{A.M. Tsvelik$^1$ and F. H. L. Essler$^2$}
\affiliation{
$^1$ Department of Condensed Matter Physics and Materials Science,
Brookhaven National Laboratory, Upton, NY 11973-5000, USA\\
$^2$ Rudolf Peierls Centre for Theoretical Physics,
University of Oxford, 1 Keble Road, Oxford OX1 3NP, UK
} 

\begin{abstract}
We consider the single particle spectral function for  a
two-dimensional clean superconductor in a regime of strong critical
thermal phase fluctuations. In the limit where the maximum of the
superconducting gap is much smaller than the Fermi energy we obtain an
exact expression for the spectral function integrated over the
momentum component perpendicular to the Fermi surface.  
\end{abstract}


\maketitle
In conventional BCS superconductors the amplitude of
the complex order parameter $|\Delta| e^{i\Phi}$ vanishes at the
transition temperature $T_c$. This is in contrast to the underdoped
cuprates, where experimental evidence \cite{Ong,exp} suggests that
the transition is instead driven by the disordering of the
superconducting phase through thermal fluctuations, while leaving the
magnitude $|\Delta|$ of the order parameter intact. 
A quantitative measure for the strength of phase fluctuations is
provided by the ratio $Q= 2T_c/\pi\rho_s(0)$, where $\rho_s(0)$ is the
zero temperature phase stiffness. This ratio determines how close
the transition is to being  mean-field-like. In BCS superconductors 
$Q\ll 1$, while in the underdoped cuprates $Q \sim 1$ \cite{EK}. 
The effects of thermal phase fluctuations on d-wave superconductors
have been investigated before, see e.g. Refs 
\onlinecite{Millis,Dorsey,Berg,khodas1,Benfatto,Eckl,Franz}. A key
objective of these works is to identify clear signatures of thermal
phase fluctuations in single particle properties such as the spectral
function measured by ARPES and STS (Scanning Tunneling
Spectroscopy). The purpose of this note is to provide an exact result
for the partially integrated spectral function of a phase fluctuating
superconductor in a particular limit. The latter quantity is defined as  
\be
\rho_P(\omega, k_{\perp}) = \int \rd k_{\parallel} A(\omega, {\bf k})\label{PDOS}
\ee
where $k_{\parallel}, k_{\perp}$ are the wave vector components
parallel and perpendicular to the Fermi velocity at the point of
observation.  
We start by summarizing the essential assumptions underlying the model
proposed by one of the authors and M. Khodas in \cite{khodas1,khodas2}. 
The starting point is a superconductor with a general order parameter
that arises from pairing on a Fermi surface the shape of which we keep
general for now. In particular it could be open or consist of several
pockets, as is believed to be the case in underdoped cuprates
\cite{Gap,RiceJ,deHaas1,deHaas}. Our following analysis is based
the existence of well defined quasiparticles, which is a reasonable
assumption for the nodal regions.
The corresponding Bogoliubov-deGennes Hamiltonian is
\bea
H &=&\int{\bf dr}\ {\bf dr'}\
\left[{\Psi^+}({\bf r})\right]^T
\Big\{\delta({\bf r} - {\bf r}')\hat\epsilon(-\ri\nabla)\tau^3 \nonumber\\
&&\qquad\qquad\qquad
+ \frac{1}{2}\widetilde{\Delta}({\bf r},{\bf r'})\tau^+
+{\rm h.c.}\Big\} \Psi({\bf r'}),
 \label{H}
\eea
where we have defined Nambu spinors $\Psi^T = (\psi_{\uparrow},
\bar\psi_{\downarrow})$, $\tau^a$ are Pauli matrices and the pairing
amplitude can be cast in the form
\be
\widetilde{\Delta}({\bf r},{\bf r'}) = \Delta({\bf r} - {\bf r}')\
\re^{\ri\phi({\bf R})}. \label{amp}
\ee
Here $\Delta({\bf r})$ determines the symmetry of the order parameter
and ${\bf R}=({\bf r}+{\bf r'})/2$ is the centre of mass co-ordinate.
Following the standard assumptions we neglect quantum fluctuations of
$\widetilde{\Delta}$ and focus exclusively on thermal fluctuations of
the phase $\phi$. The key point is to choose an appropriate model for
these phase fluctuations. The effects of fully three dimensional
fluctuations are well studied in the literature \cite{larkin} and are
found to be small. On the other hand, one would expect the spatial
anisotropy of layered materials like the cuprates to strongly enhance
the role of phase fluctuations. The extreme limit would be the purely
two dimensional case, on which we focus in what follows. We
emphasize that even purely 2D models have a window of applicability to
e.g. thin films \cite{Bosovic} and $x=1/8$ LBCO, where the phase
transition was found to be of Berezinskii-Kosterlitz-Thouless (BKT) type
\cite{Tranq},\cite{Basov},\cite{Tranq1}. Similarly, the analyses of the temperature dependence of
magnetization, London penetration \cite{Cam} depth and terahertz conductivity \cite{orenstein} for high quality underdoped
BiSCO crystals show that although the superconductivity below $T_c$ is
of a 3D nature, the superconducting transition in these systems is
rather close to a BKT transition. In the latter  case our
theory will be applicable in a temperature regime above $T_c$, where the
phase correlation length is exponentially large and the phase
fluctuations can effectively be considered as critical.        
  
In the mean field approximation fluctuations of the order parameter
$\widetilde{\Delta}$  are ignored, and the resulting Green function
takes the familiar BSC form
\be
G_{\rm BCS}(\omega,{\bf k}) = \frac{\omega + \epsilon_{\bf k}}
{(\omega + \ri 0)^{2} -\epsilon_{\bf k}^{2} - \Delta^{2}({\bf k})}.
\ee
The corresponding spectral function $-\frac{1}{\pi}{\rm Im}\ G_{\rm BCS}$
consists of two delta function peaks centered at positive
and negative frequencies. These peaks will be broadened by thermal phase
fluctuations. The following facts are of crucial importance
in the following: (i) Since the long wavelength
fluctuations are classical, the electron frequency is conserved. 
(ii) Since in the region of interest the amplitude $|\Delta|$ is
assumed to be fixed, self-consistency between the electron Green
function and the order parameter is not an issue. Hence the calculation
of the spectral function is reduced to solving the Bogoliubov-deGennes
equations for a particle with pairing amplitude (\ref{amp}) and then
averaging the result over a given distribution of phase fluctuations.
(iii) Since we are interested only in long wavelength
fluctuations, the distribution function $P(\phi)=\re^{-F_{\phi}/T}$ can be
fixed by symmetry considerations: as long as the discrete lattice
symmetries include $C_{4}$, the group of in-plane rotations by ninety
degrees, the distribution function must be spatially isotropic (apart
from irrelevant higher gradient terms). This leads to 
\begin{align}\label{2d}
\frac{F_{\phi}}{T} = \frac{\rho_s}{2T}\int \rd x \rd y \left[ (\p_{x}\phi)^2
  + (\p_{y}\phi)^2  \right] \, ,  
\end{align}
where the prefactor $T^{-1}$ results from the integration over
imaginary time. In contrast to the phase fluctuations the Green's
function at low energies is very different in the directions
perpendicular and tangential to the Fermi surface. In a process where an
electron close to the Fermi surface changes its momentum from
${\bf k}$ to ${\bf k} + {\bf q}$ by scattering off the pairing
potential its Green's function is
\bea
G_0^{-1}(\omega,{\bf k}+{\bf q}) 
&=& \omega - \epsilon({\bf k} + {\bf q})\nonumber\\
& \approx& \omega - \epsilon({\bf k}) - vq_{\parallel} -
\frac{q_{\perp}^2}{2m}. \label{G0} 
\eea
Here $q_{\parallel}$ and $q_{\perp}$ are the components of the
momentum respectively parallel and perpendicular to the Fermi velocity
$\nabla\epsilon({\bf k})$. As a result of the isotropy of the
distribution $F_{\phi}$ of phase fluctuations the typical values of
$q_{\parallel}$ and $q_{\perp}$ are the same and of order $\Delta_{\rm
  max}$ (the maximal gap). Therefore the last term in  \eqref{G0} is
proportional to the small parameter $\Delta/\epsilon_F$. 
If we neglect such small corrections the electron propagates along a
straight line in real space and the transverse momentum is conserved.  
The electron Green's function can then be calculated separately for
each frequency $\omega$ and Fermi surface point ${\bf k}$.
\begin{figure}[ht]
\epsfxsize=0.3\textwidth
\epsfbox{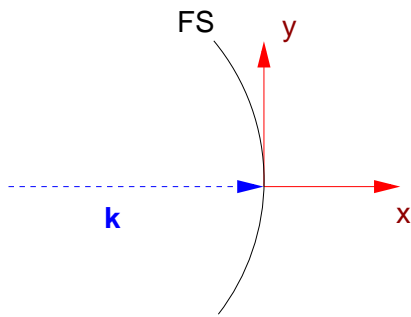}
\caption{Geometry defining the model in (\ref{action}).}
\label{fig:fs}
\end{figure}
Under the assumptions summarized above the initial  problem (\ref{H})
is recast as a field theory described by the Lagrangian
$\mathcal{L}= F_{\phi}  + \bar{\Psi}_{\omega_n} {\cal H}\
\Psi_{\omega_n}$ with
\bea
\label{action}
{\cal H}&=& 
 - \ri \omega_{n} I - \ri v\tau^{z}  \partial_{x} 
+\frac{\widetilde{\Delta}(k_{\perp},x)}{2} \tau^{+} \!+
\frac{\widetilde{\Delta}^{*}(k_{\perp},x)}{2} \tau^{-},\nonumber
\eea
where $\widetilde{\Delta}(k_{\perp},x) =
\Delta(k_{\perp})\re^{\ri\phi(x,0)}$ and we have introduced
$\Psi_{\omega_n}=(\psi_{\omega_n,\uparrow}, \psi^\dagger_{-\omega_n,\downarrow})^T$.
In (\ref{action}) we have used co-ordinates as shown in
Fig.\ref{fig:fs}. As was pointed out in Ref.\cite{khodas1}, the model
defined through Eqs~\eqref{action} and \eqref{2d} is in fact
equivalent to the anisotropic spin-1/2 Kondo problem. In terms of this
impurity model the phase fluctuations play the role of the host, while
a \emph{single} Bogoliubov quasiparticle constitutes the magnetic
impurity. The reduction of the underlying interacting electron model
to a single-impurity problem is possible because the emergent
low-energy degrees of freedom are non-interacting Bogoliubov
quasiparticles. The many-body aspects of the problem enter the
determination of $|\Delta|$, but as this is treated as a parameter of
our model we can avoid the issue of its calculation. Under a field
redefinition  
\be
\begin{pmatrix}
\psi_{\omega_n,\uparrow}\\ \psi^{\dagger}_{-\omega_n,\downarrow}
\end{pmatrix}=
\begin{pmatrix}
\chi_{\omega_n,\uparrow}\\ -i\chi^{\dagger}_{-\omega_n,\downarrow}
\end{pmatrix}\ ,\ \
\begin{pmatrix}
\psi^\dagger_{\omega_n,\uparrow}\\ \psi_{-\omega_n,\downarrow}
\end{pmatrix}=
\begin{pmatrix}
i\chi_{\omega_n,\uparrow}\\ \chi^{\dagger}_{-\omega_n,\downarrow}
\end{pmatrix},
\label{newv}
\ee
and subsequent analytic continuation $\ri \omega_{n} \rightarrow
\omega + \ri 0$ we obtain the Hamiltonian
\bea
H_{\rm eff} &=&  v^{-1}\ri(\omega +\ri 0)\hat\tau^3 
+ H_{\rm bulk}[\phi]  \label{Leggett}\\
&+& \frac{\Delta(k_{\perp})}{2v}\Big[
\hat\tau^+\re^{\ri\phi(y=0)} + \hat\tau^-\re^{-\ri\phi(y=0)}\Big], \nonumber
\eea
where $\hat\tau^a \equiv \chi^+\tau^a\chi$ is a short hand notation
for fermionic bilinears.
In this setting the coordinate $x$ plays the role of Matsubara
time. It is dual to the momentum component $k_{\parallel}$ parallel to
the Fermi velocity at the point of observation. We note that in the
approximation underlying (\ref{action}) the electron momentum parallel
to the Fermi surface is conserved so that the fermions $\chi$  
depend only on $x$, while the phase field $\phi$ is a function of both
$x$ and $y$. For convenience we assign $\chi$ the coordinate 
$y=0$. Since the fermion number is conserved, the
$\hat{\tau}$-operators are in fact components of a spin S=1/2. 

The Hamiltonian $H_{\rm bulk}$ arising from (\ref{2d}) describes the
phase fluctuations. For temperatures below the
BKT transition temperature 
$T_{\mathrm{BKT}} = \pi \rho_{s}/2$ only smooth field configurations
contribute so that 
\bea
 H_{\rm bulk}[\phi]= \frac{1}{8\pi d}\int_{-\infty}^{\infty} \rd y 
 \left[  (4 \pi d)^{2}  \Pi^2 + (\p_y\phi)^2 \right], \label{Gauss}
\eea
where $\Pi$ is the momentum density conjugate to the field $\phi$,
with equal time commutator 
$\left[ \Pi(y_{1}),\phi(y_{2}) \right]_{-} = - \ri \delta(y_{1} - y_{2})$.
In order to be able to treat the temperature region $T > T_{\rm BKT}$
we need to allow singular (vortex) configurations of the $\phi$
field. The effects of vortices can be illustrated for the example
of the two point correlation function of bosonic exponents . The
latter takes the form
\bea
\la \re^{\ri \phi({\bf r}_1)}\re^{-\ri\phi({\bf r}_2)}\ra = \left| \frac{b}{\xi(T)}\right|^{2d}F\left(\frac{{\bf r}_{12}}{\xi(T)}\right) \, , \label{2point}
\eea
where $d = T/(8T_{\mathrm{BKT}})$ is the scaling dimension of the
order parameter, $\xi(T)$ is the correlation length
and $b \sim (v/\epsilon_{F})$ is the short distance cut-off. 
The short and long distance behaviour of the scaling function is 
$F(\rho\ll 1) =\rho^{-2d}$ and $F(\rho > 1) \sim K_0(\rho)$ respectively
(see also \cite{LukZam}). 
Below the transition (where $\xi = \infty$) the function
(\ref{2point}) decays as a power law and above the transition where the
vortices are relevant it decays exponentially with finite correlation
length $\xi(T)$. We show below how to take this into account.
It was shown in Ref \cite{schotte} that \eqref{Leggett},
\eqref{Gauss}  is equivalent to the anisotropic S=1/2 Kondo model in
the regime of extreme anisotropy $g_{\parallel} \gg g_{\perp}$
\bea
&&H_{\rm Kondo} = 
\sum_k vk a^+_{k\s}a_{k\s} + h\tau^z
\label{Kondo}\\
&&\quad+  \frac{J}{N}\sum_{p,k}g_\parallel a^+_{k\s}\tau^z_{\s\s'}a_{p\s'} + 
\frac{g_{\perp}}{2}
\Big[a^+_{k\s}\tau^+_{\s\s'}a_{p\s'} + h.c.\Big],\nonumber
\eea
where the magnetic field $h$ is related to the real frequency $\omega$
in (\ref{Leggett}) by analytic continuation $h=i\omega+0$.
Our main result derives from the observation that the
partial density of states (PDOS) defined by \eqref{PDOS} is equal to the 
Green function of $\chi$ fermions at coinciding coordinates $x$. 
Taking into account the change of variables \eqref{newv} 
we find that the PDOS is obtained by analytic continuation of the
impurity magnetization of the Kondo model \eqref{Kondo}
\bea
\rho_P(\omega)/\rho_0 = 2{\rm Re}\ M(h = \ri\omega +0)\ .
\label{rho}
\eea
Here $\rho_0$ is the bare density of states. This expression provides
a link between spectral properties of the single electron problem
\eqref{H} and thermodynamic properties of the many-body theory
\eqref{Kondo}. 
In order to utilize the known exact expression for $M(h)$ in the Kondo
problem \cite{wiegmann81} we need to relate the parameters
$Jg_\parallel$, $Jg_\perp$ in (\ref{Kondo}) to $d$ and $\Delta(k_\perp)$.
The interactions in the Kondo model increase under renormalization
and enter the strong coupling regime at a scale $T_H$ which is known
from the exact solution \cite{wiegmann81}
\be
T_H \sim \epsilon_F (g_{\perp}/g_{\parallel})^{2\pi/g_{\parallel}}.
\ee
On the other hand the usual scaling argument gives
$T_H \sim \epsilon_F  g_{\perp}^{1/(1-d)}$,
which leads to the identification 
$g_{\parallel}/2\pi =1-d$ with $d = T/(8T_{\rm BKT})$. 
The expression for the impurity magnetization derived in
\cite{wiegmann81} then reads (the parameter $\mu$ in \cite{wiegmann81} is related to $d$ by $\mu=\pi(1-d)$): 
\bea
M(h/T_H) &=& \frac{\ri}{4\pi^{3/2}}\int_{-\infty}^{\infty}\frac{\rd x}{x
  + \ri 0}
\frac{\Gamma\Big(1- \ri\frac{x}{1-d}\Big)\Gamma\Big(\frac{1}{2} + \ri
  x\Big)}{\Gamma\Big(1- \ri \frac{x d}{1-d}\Big)}\nonumber\\
&&\times\ \exp\Big\{-2\ri x\Big[\ln(h/T_H) +\pi
a\Big]\Big\} \label{M},
\eea
where $\pi a = \frac{1}{2(1-d)}[d\ln d +(1-d)\ln(1-d)]$.
As a function of a complex variable $M(z)$ 
admits a power series expansion in odd powers of $z$ for $|z| <1$
and concomitantly is purely imaginary along the imaginary axis. 
By virtue of the identification (\ref{rho}) this implies that the PDOS
vanishes at $|\omega| < T_H$. \emph{Thus there is a sharp gap equal to
  $T_H$ in the density of states}, which at $T \neq 0$ is always
smaller than the mean field gap $\Delta(k_\perp)$. On the other hand,
for $|z| > 1$ the following expansion holds 
\bea
\label{Pser}
\frac{\rho_P(\omega)}{\rho_0} &=& 1 +
\sum_{n=1}^{\infty} \frac{\sin[2\pi
  nd]}{2\pi^{3/2}(n!)}\Gamma(nd)\Gamma\Big(\frac{1}{2} +
(1-d)n\Big)\nonumber\\
&&\times
\Big[\frac{T_H\re^{-\pi a}}{\omega}\Big]^{2n(1-d)}, ~~
|\omega| > T_H.
\eea
We note that
\be
\lim_{d\to 0}\frac{\rho_P(\omega)}{\rho_0} =
\frac{|\omega|}{\sqrt{\omega^2-\Delta^2}}\theta(|\omega|-\Delta)\ ,
\ee
which corresponds to the mean field result. In order to establish the
relation between the gap $T_H$ and $\Delta(k_\perp)$, $d$ we compare
\eqref{Pser} to the perturbative expansion for the PDOS in the model
(\ref{Leggett}). Second order perturbation theory gives
 \bea
\frac{\delta\rho_P}{\rho_0} 
 &=& \frac{\cos(\pi d)\Gamma(2-2d) 2^{2d}\Delta^2
   b^{2d}}{2\omega^{2(1-d)}},
\label{corr2} 
\eea
which yields the desired identification
\be
 T_H = \Delta(k_{\perp})\sqrt{1-d}\Big[\sqrt d
 b\Delta(k_{\perp})\Big]^{d/1-d}\Big[\Gamma(1-d)\Big]^{1/1-d}.
 \label{TH} 
\ee
\begin{figure}[ht]
\epsfxsize=0.45\textwidth
\epsfbox{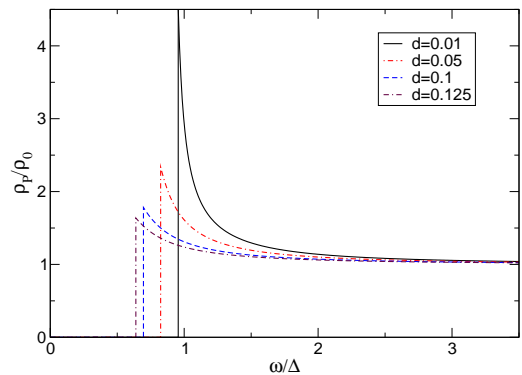}
\caption{Partial DOS as a function of frequency plotted for different
temperatures $T=8d\ T_{\rm BKT}$. The ratio of $\Delta(k_\perp)$ to
the cutoff $1/b$ is fixed as $0.1$. Due to particle-hole symmetry
$\rho_P(-\omega)=\rho_P(\omega)$.}
 \label{fig:FS}
\end{figure}
Given the result (\ref{Pser}) for the PDOS we may calculate the full
tunneling density of states. In the case of a d-wave superconductor
this results in
\be
\rho(\omega) \propto |b\omega|^{1-d}.
 \label{DOS}
\ee
In Fig.~\ref{fig:FS} we show the PDOS (\ref{Pser}) as a function of
frequency for several different temperatures. The most noticeable
feature is the persistence of a sharp gap. In addition we observe that
the singularity characteristic of the BCS mean-field solution is
strongly suppressed as $T$ increases. This demonstrates that thermal phase
fluctuations have a sizable effect on integrated spectral properties. 
In realistic materials the sharp gap will be smeared by both impurity
scattering and the effects of Fermi surface curvature neglected in our
analysis. 
\begin{figure}[ht]
\epsfxsize=0.45\textwidth
\epsfbox{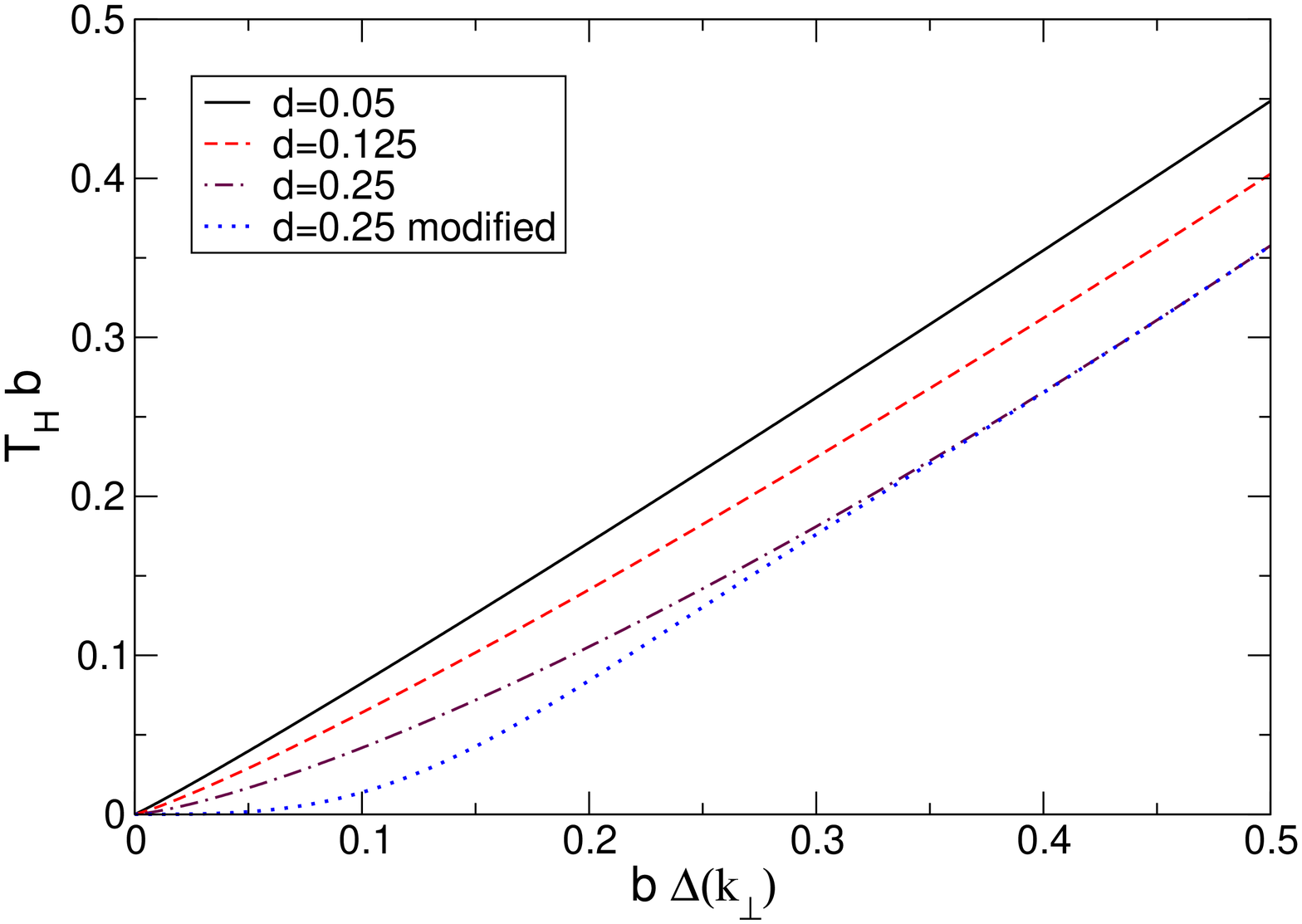}
\caption{The dimensionless gap $T_H\ b$ as a function of $q
  =\Delta(k_{\perp})b$ for $d=0.05, 0.125, 0. 25$. The lowest curve,
corresponding to $T = 2T_{BKT}$, has been modified to indicate the
effects of vortices as described in the main text.  
 \label{fig:TH} }
\end{figure}
In Fig.~\ref{fig:TH} we plot the temperature evolution of the gap
$T_H$. We see that temperature effects are negligible.
In particular this implies that the ($d$-wave) form of
the gap remains robust through the transition, at least if
vortices are ignored. As discussed above, the main effect of vortices
is to induce a finite correlation length $\xi(T)$. In the
corresponding Kondo picture this translates to a gap in the excitation
spectrum of the host. Exact results are available in this case as well
\cite{schlottmann}. On a qualitative level what happens in the
Kondo picture is the following: as long as the correlation length
$\xi(T)$ is larger than the inverse Kondo scale $v/T_H$ the vortices 
have little effect on the physical properties. However, as soon as
$\xi(T)$ falls below $v/T_H$ the scaling terminates before the strong
coupling regime is reached. As a consequence the gap in the PDOS is
reduced for momenta close to the node $k_{\perp} < 1/[b\Delta
\xi(T)]$. We have indicated this effect in the dotted curve in
Fig.\ref{fig:TH}. 

In this work we have considered a model for thermal phase fluctuations
in a superconductor recently proposed in Ref.~\cite{khodas1}. By
exploiting a mapping to an effective spin-1/2 Kondo problem we have
derived an \emph{exact} result for the partially integrated spectral
function (\ref{PDOS}). Our main result is that thermal fluctuations
have a substantial effect on the single particle spectral function.
The best candidate for comparing our theory to experiment is x=1/8
doped LBCO. ARPES and STS measurements performed in \cite{valla} show
that the $d$-wave gap is already well formed at the BKT
transition. It would be interesting to map out the detailed
temperature dependence of the spectral function by ARPES in the region
of strong diamagnetic fluctuations $T < 40$K and carry out a partial
integration along the nodal direction.

We thank M. Khodas for important discussions.
This work was supported by the Center for Emerging Superconductivity
funded by the U.S. Department of Energy, Office of Science (AMT) by
the EPSRC under grants EP/D050952/1 (FHLE) and EP/H021639/1 (AMT and
FHLE).

\end{document}